\documentclass[epj]{webofc}
\usepackage[varg]{txfonts}   
\usepackage{amsmath,amssymb,amstext}
\usepackage{siunitx}
\usepackage{color}
\usepackage{graphicx}

\newcommand{\dd}{\ensuremath{\mathrm{d}}}

\woctitle{MESON2016 - the 14$^\textrm{th}$ International Workshop on Meson Production, Properties and Interaction}
\begin{document}
\selectlanguage{english}
\title{Double polarisation experiments in meson photoproduction}

\author{Jan~Hartmann\inst{1}\fnsep\thanks{\email{hartmann@hiskp.uni-bonn.de}}
       \\for the CBELSA/TAPS Collaboration
}

\institute{Helmholtz-Institut f\"ur Strahlen- und Kernphysik, Universit\"at Bonn, Germany}

\abstract{
One of the remaining challenges within the standard model is to gain a good understanding of QCD in the non-perturbative regime. A key step towards this aim is baryon spectroscopy, investigating the spectrum and the properties of baryon resonances. To gain access to resonances with small $\pi N$ partial width, photoproduction experiments provide essential information. Partial wave analyses need to be performed to extract the contributing resonances. Here, a complete experiment is required to unambiguously determine the contributing amplitudes. This involves the measurement of carefully chosen single and double polarisation observables. 
In a joint endeavour by MAMI, ELSA, and Jefferson Laboratory, a new generation of experiments with polarised beams, polarised proton and neutron targets, and $4\pi$ particle detectors have been performed in recent years. Many results of unprecedented quality were recently published by all three experiments, and included by the various partial wave analysis groups in their analyses, leading to substantial improvements, e.g. a more precise determination of resonance parameters.
An overview of recent results is given, with an emphasis on results from the CBELSA/TAPS experiment, and their impact on our understanding of the nucleon excitation spectrum is discussed.
}
\maketitle
\section{Introduction}
The spectrum of excited nucleon states reflects the dynamics of QCD in the non-perturbative regime. It has been studied for many years using pion beams. However, the spectrum of known nucleon resonances is in conflict with predictions from quark models \cite{isgur77,loring01}. Most obvious is the missing resonance problem: the fact that more states are predicted by  models at higher masses than have been observed experimentally. But also the ordering of excited states with positive and negative parity is partly in disagreement with experiments, the most prominent example being the $N(1440)\,1/2^+$ which is predicted by most quark models to be heavier than the $N(1535)\,1/2^-$. QCD calculations on the lattice \cite{edwards11}, though using unphysically large quark masses, yield a similar pattern as the non-relativistic quark model. Measuring the properties of the known resonances more precisely and searching for new resonances is essential to understanding the discrepancies between theory and experiment.

Photoproduction experiments allow access to resonances with small $\pi N$ couplings and therefore have great potential to observe the missing resonances. The contributing resonances are extracted from the measured data in a partial wave analysis (PWA). To determine the contributing amplitudes in an unambiguous way, a complete experiment \cite{chiang97} is needed, which requires the measurement of polarisation observables.
In this paper, a selection of recent measurements of single and double polarisation observables accessible with linearly or circulary polarised photon beam and a longitudinally or transversely polarised proton target is shown, and their impact on the PWA is discussed. 

\section{Experimental setup}
The data presented here were obtained with the CBELSA/TAPS experiment at ELSA \cite{hillert06}. A linearly polarised photon beam was produced from the incident $\SI{3.2}{GeV}$ electron beam via coherent bremsstrahlung off a carefully aligned diamond crystal \cite{elsner09}. For a coherent edge at $E_{\gamma} = \SI{950}{MeV}$ a maximum polarisation of $\SI{65}{\percent}$ was reached. A circularly polarised photon beam was produced from a longitudinally polarised $\SI{2.4}{GeV}$ electron beam provided by ELSA. The polarisation was measured with a M\o ller polarimeter, a maximum electron polarisation of $\SI{64}{\percent}$ was reached.
The electrons, after hitting the radiator target, passed through a magnet onto a tagging hodoscope which defined the energy of the bremsstrahlung photons. The photon beam impinged on a frozen spin butanol target \cite{bradtke99} providing transversely polarised protons with an average target polarisation degree of $\SI{74}\%$.

\begin{figure}[b]
	\centering
	\includegraphics[width=.9\textwidth]{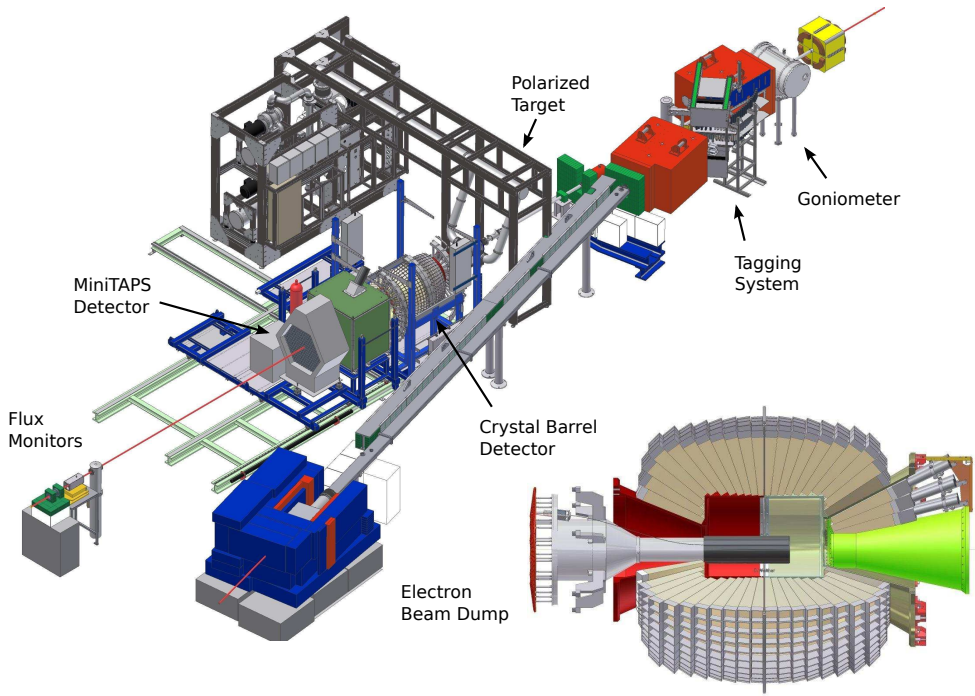}
	\caption{The experimental setup of the CBELSA/TAPS experiment.}
	\label{fig:setup}
\end{figure}
The detector system, which is shown in Figure~\ref{fig:setup}, consisted of two electromagnetic calorimeters, the Crystal Barrel \cite{aker92} and the MiniTAPS detector \cite{novotny91}, together covering the polar angle range from $1^\circ$ to $156^\circ$ and the full azimuthal angle. For charged particle identification, a three-layer scintillating fibre detector \cite{suft05} surrounding the target, and plastic scintillators in forward direction could be used. The detector setup provided a high detection efficiency for photons and is therefore ideally suited to measure single and double polarisation observables in reactions with neutral mesons decaying into photons in the final state.

\section{Results}
The cross-section for photoproduction of single pseudoscaler mesons with a polarised beam off a polarised target can be written in the form \cite{barker75}
\begin{align}
	\label{eq:xsect} 
	\frac{\dd\sigma}{\dd\varOmega} = \left(\frac{\dd\sigma}{\dd\varOmega}\right)_{0} \cdot \Big[
		& {1} - {\varSigma}{\delta_\ell\cos(2\phi)} + {T}{\varLambda_y} - {P}{\varLambda_y\delta_\ell\cos(2\phi)} \\
		& - {E}{\varLambda_z\delta_\odot} + {F}{\varLambda_x\delta_\odot} + {G}{\varLambda_z\delta_\ell\sin(2\phi)} + {H}{\varLambda_x\delta_\ell\sin(2\phi)} \Big], \nonumber
\end{align}
where $\left(\frac{\dd\sigma}{\dd\varOmega}\right)_0$ is the unpolarised cross section, $\delta_\ell$ ($\delta_\odot$) denotes the linear (circular) polarisation degree of the photon beam, and $\varLambda_{x,y,z}$ is the polarisation degree of the target protons in the centre-of-mass (CM) coordinate frame. In a polarised butanol (C$_4$H$_9$OH) target, not only the polarised protons, but also unpolarised nucleons in the C and O nuclei contribute to the count rate. This effectively reduces the target polarisation and can be taken into account by the dilution factor $d$: $\varLambda \to d\,\varLambda$. To determine $d$, the background contribution from bound nucleons needs to be determined, e.g.\@ using a background measurement with a carbon target. For details, see e.g.\@ \cite{gottschall14,hartmann15}.

\subsection{\boldmath $\pi^0$ photoproduction}
\begin{figure}[b]
	\centering%
	\hfill%
	\includegraphics[width=.4\textwidth]{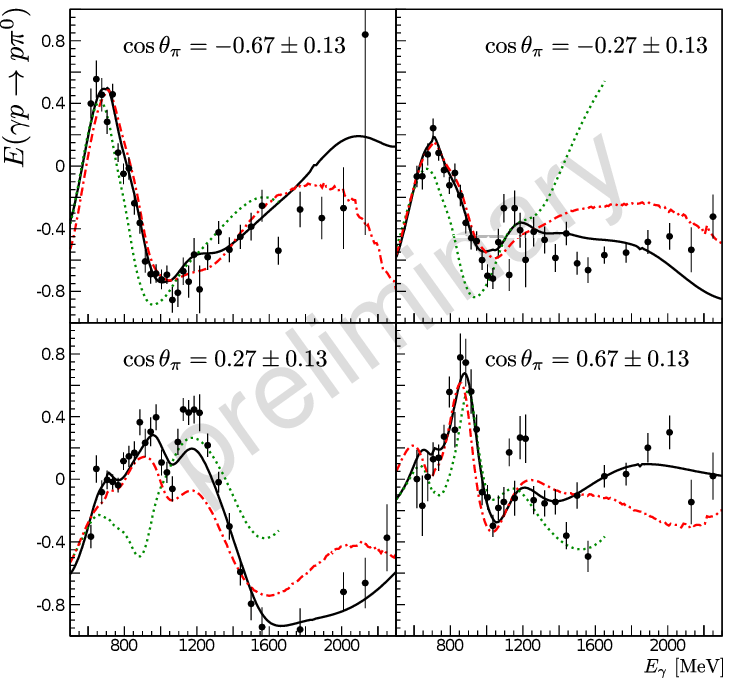}%
	\hfill%
	\includegraphics[width=.4\textwidth]{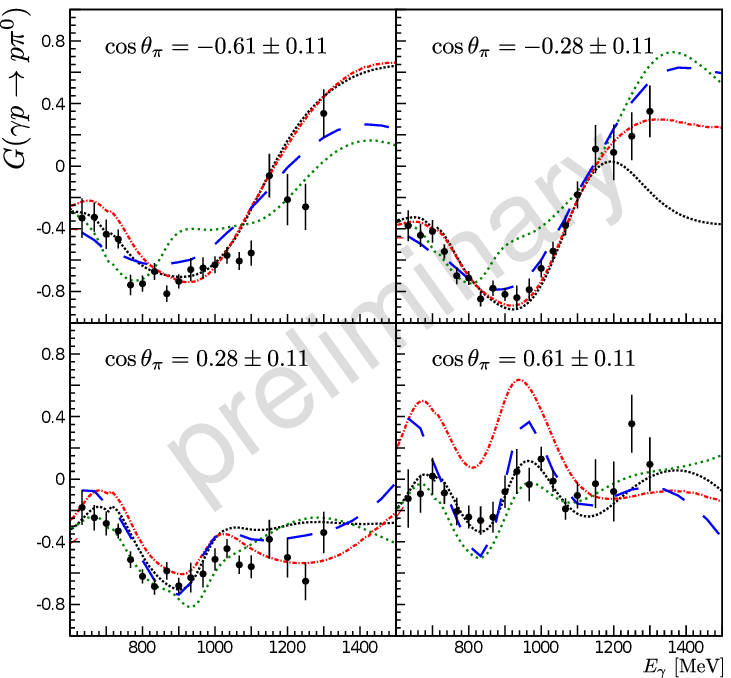}%
	\hfill%
	\vspace*{-2mm}%
	\caption{Results for the observables $E$ (left) \cite{gottschall14} and $G$ (right) \cite{thiel12,thiel16} for four different angular bins, compared to different PWA predictions: MAID (green, dotted) \cite{maid07}, SAID CM12 (red, dashed-dotted) \cite{said12}, BnGa 2011-02 (black, solid) \cite{anisovich12}, and J\"uBo 13-01 (blue, dashed) \cite{roenchen14}.}%
	\label{fig:pi0_EG}%
\end{figure}
With circularly polarised photons and longitudinally polarised protons, the helicity asymmetry $E$ is accessible. It can be determined from the asymmetry of events with parallel ($N_{3/2}$) and anti-parallell ($N_{1/2}$) spins of the initial-state particles as
\begin{equation}
	E = \frac{N_{1/2} - N_{3/2}}{N_{1/2} + N_{3/2}} \cdot \frac{1}{d\,\varLambda_z\,\delta_\odot}.
\end{equation}
With linearly polarised photons and logitudinally polarised protons, the double polarisation observable $G$ can be deduced from the correlation between the photon polarisation plane and the scattering plane. The number of events $N$ as a function of the azimuthal angle $\phi$ between the two planes is given by
\begin{equation}
	N(\phi) / N_0 = 1 - \delta_\ell\,\varSigma_\mathrm{eff} \cos(2\phi) + d\,\varLambda_z\,\delta_\ell\,G \sin(2\phi),
\end{equation}
where $N_0$ is given by averaging $N(\phi)$ over $\phi$. $\varSigma_\mathrm{eff}$ mixes the beam asymmetry from free and bound nucleons.
Results for both $E$ and $G$ are shown in Fig.~\ref{fig:pi0_EG} in comparison to different PWA predictions.

\begin{figure}[t]
	\centering%
	\hfill%
	\includegraphics[height=7.5cm]{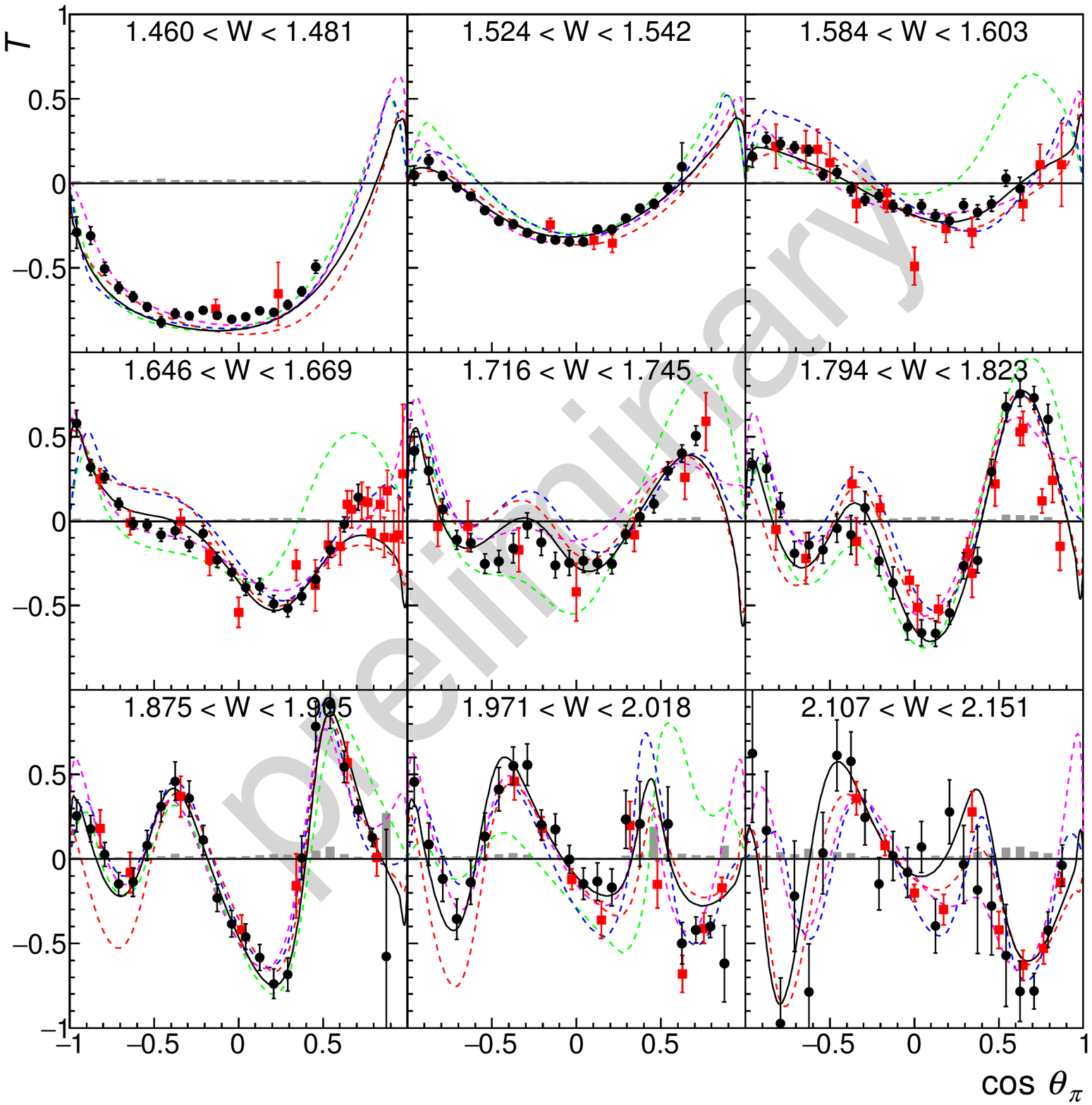}%
	\hfill%
	\includegraphics[height=7.5cm]{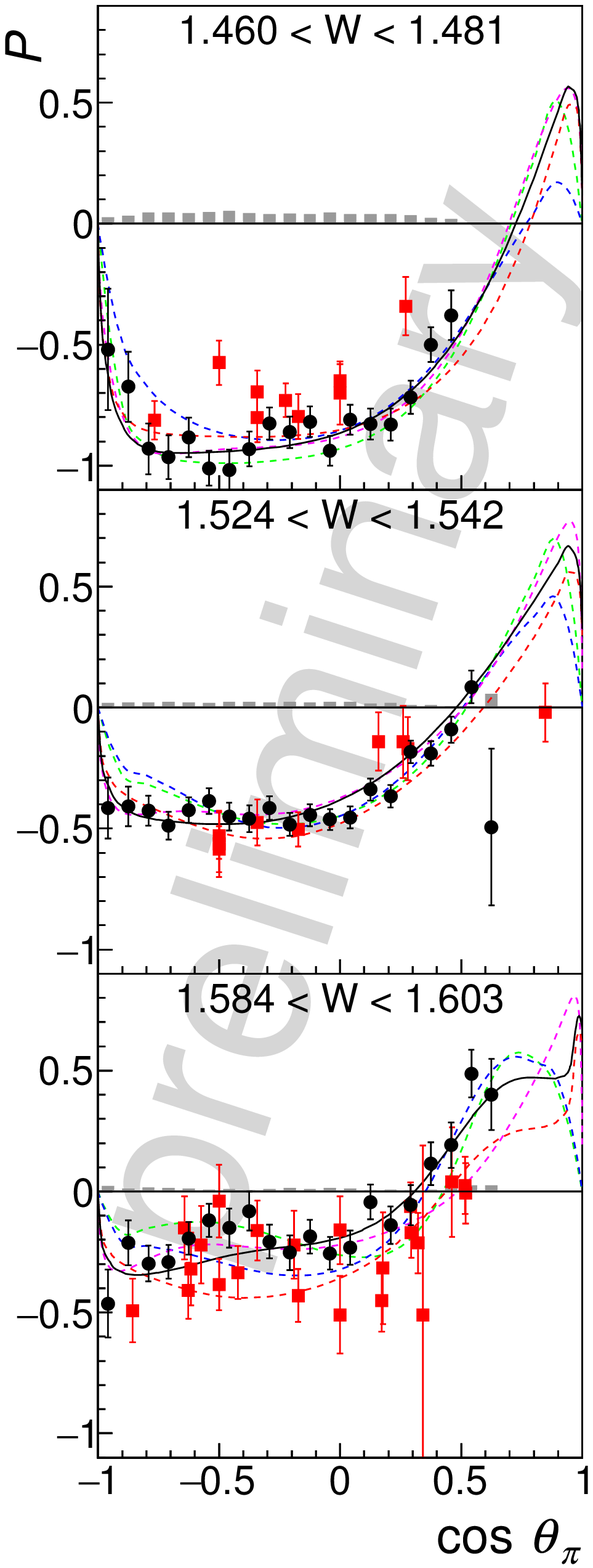}%
	\hfill%
	\includegraphics[height=7.5cm]{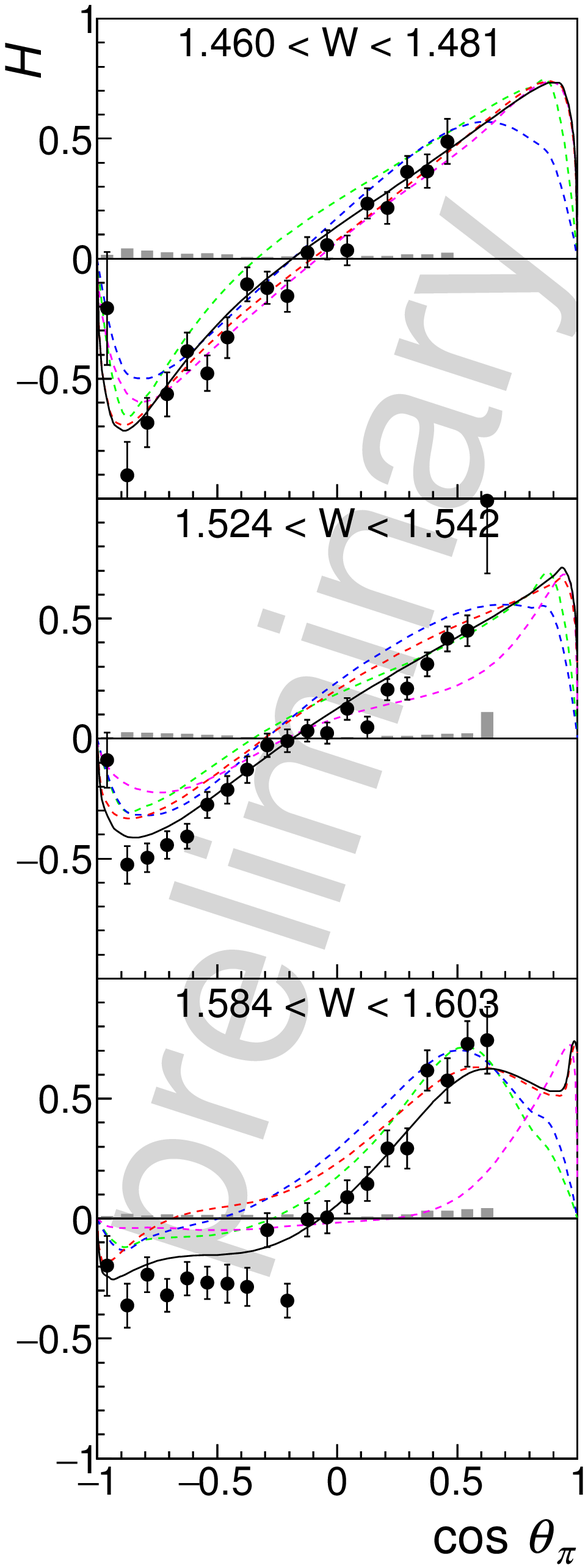}%
	\hfill%
	\vspace*{-2mm}%
	\caption{Results for the observables $T$, $P$, and $H$ \cite{hartmann14,hartmann15} as a function of the scattering angle and the $\gamma p$ invariant mass W (in GeV, only every third bin is shown here). References to earlier data (red points) are given in \cite{anisovich12}, Refs. [49-71] therein. The solid black line represents the BnGa2014 fit \cite{hartmann15}. The data are compared to predictions (dashed curves) from BnGa2011-02 (red) \cite{anisovich12}, MAID (green) \cite{maid07}, SAID CM12 (blue) \cite{said12}, and J\"uBo 2015 (magenta) \cite{roenchen15}.}%
	\label{fig:pi0_TPH}%
\end{figure}

With linearly polarised photons and transversely polarised protons, the observables $T$, $P$, and $H$ can be measured simultaneously. In this case, the azimuthal distribution of events is given by
\begin{align}
	N(\phi) / N_0 = 1 &- \delta_\ell\,\varSigma_\mathrm{eff} \cos(2\phi) + d\,\varLambda_T\,T \sin(\phi-\varphi) \\
	                  &- d\,\varLambda_T\,\delta_\ell P\cos(2\phi)\sin(\phi-\varphi) + d\,\varLambda_T\,\delta_\ell H \sin(2\phi)\cos(\phi-\varphi), \nonumber
\end{align}
where $\varphi$ is the azimuthal angle between the target polarisation vector and the photon polarisation plane. Results are shown in Fig.~\ref{fig:pi0_TPH}.

\begin{figure}[t]
	\centering%
	\sidecaption
	\includegraphics[width=.5\textwidth]{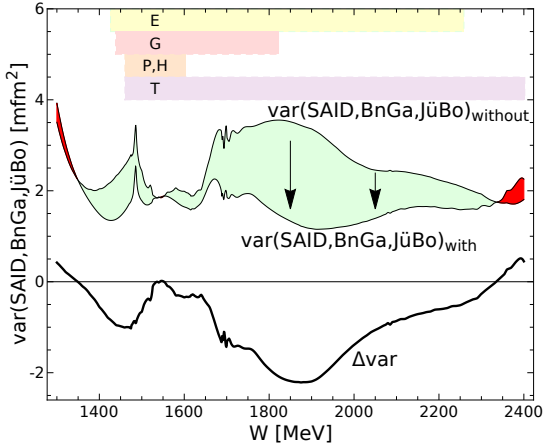}%
	\vspace*{-2mm}%
	\caption{The variance of all three PWAs summed over all $\gamma p \to \pi^0 p$ multipoles up to $L=4$ \cite{pwa16}. The range covered by the new double-polarisation observables is indicated by shaded areas. Over the largest part of the energy range the new data have enforced an improvement of the overall consistency. The improvement is displayed as a light green area and, separately, as difference of the variance. Ranges with an overall deterioration are marked in red.}%
	\label{fig:pwa_var}%
\end{figure}

Our data up to $E_{\gamma} = \SI{930}{MeV}$ were used as a basis for an energy-independent PWA \cite{hartmann14}, allowing for the determination of the $N(1520)\,3/2^-$ helicity amplitudes with minimal model dependence. 
All the data were included in the BnGa multi-channel PWA, together with further data on other channels.%
\footnote{For a complete list, see \cite{hartmann15}, Ref. [25] therein.}
Starting from the previous solutions BnGa2011-01 and BnGa2011-02 \cite{anisovich12} all parameters were re-optimised. The newly determined multipoles are compatible with the previous ones at the $2\sigma$ level over the full mass range. The errors are significantly reduced, on average by a factor of 2.25 \cite{hartmann15}. The impact of the new data on the BnGa, SAID, and J\"uBo analyses was recently investigated in a joint effort of the analysis groups \cite{pwa16}. To quantify the discrepancy between different analyses, the variance of the 16 (complex) multipole amplitudes $\mathcal{M}$ up to $L=4$ can be used, defined as
\begin{equation}
	\mathrm{var}(1,2) = \frac{1}{2} \sum\limits_{i=1}^{16} \bigl( \mathcal{M}_1(i)-\mathcal{M}_2(i)\bigr) \bigl( \mathcal{M}^*_1(i)-\mathcal{M}^*_2(i)\bigr).
\end{equation}
Once the new data are included, the discrepancy between all three analyses is significantly reduced, as can be seen in Fig.~\ref{fig:pwa_var}.

\subsection{\boldmath $\eta$ photoproduction}
\begin{figure}[b]
	\centering%
	\hfill%
	\includegraphics[height=5.2cm]{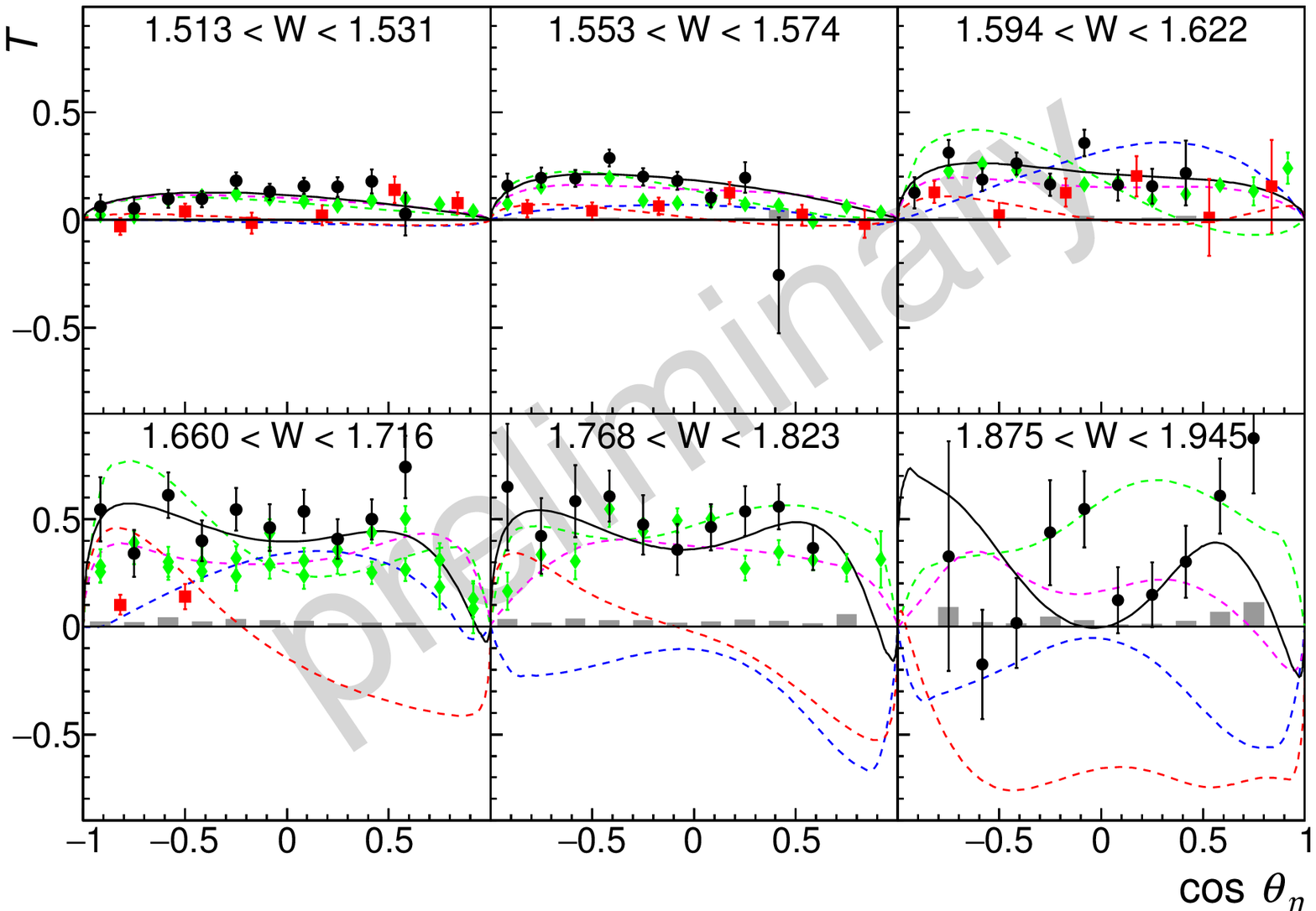}%
	\hfill%
	\includegraphics[height=5.2cm]{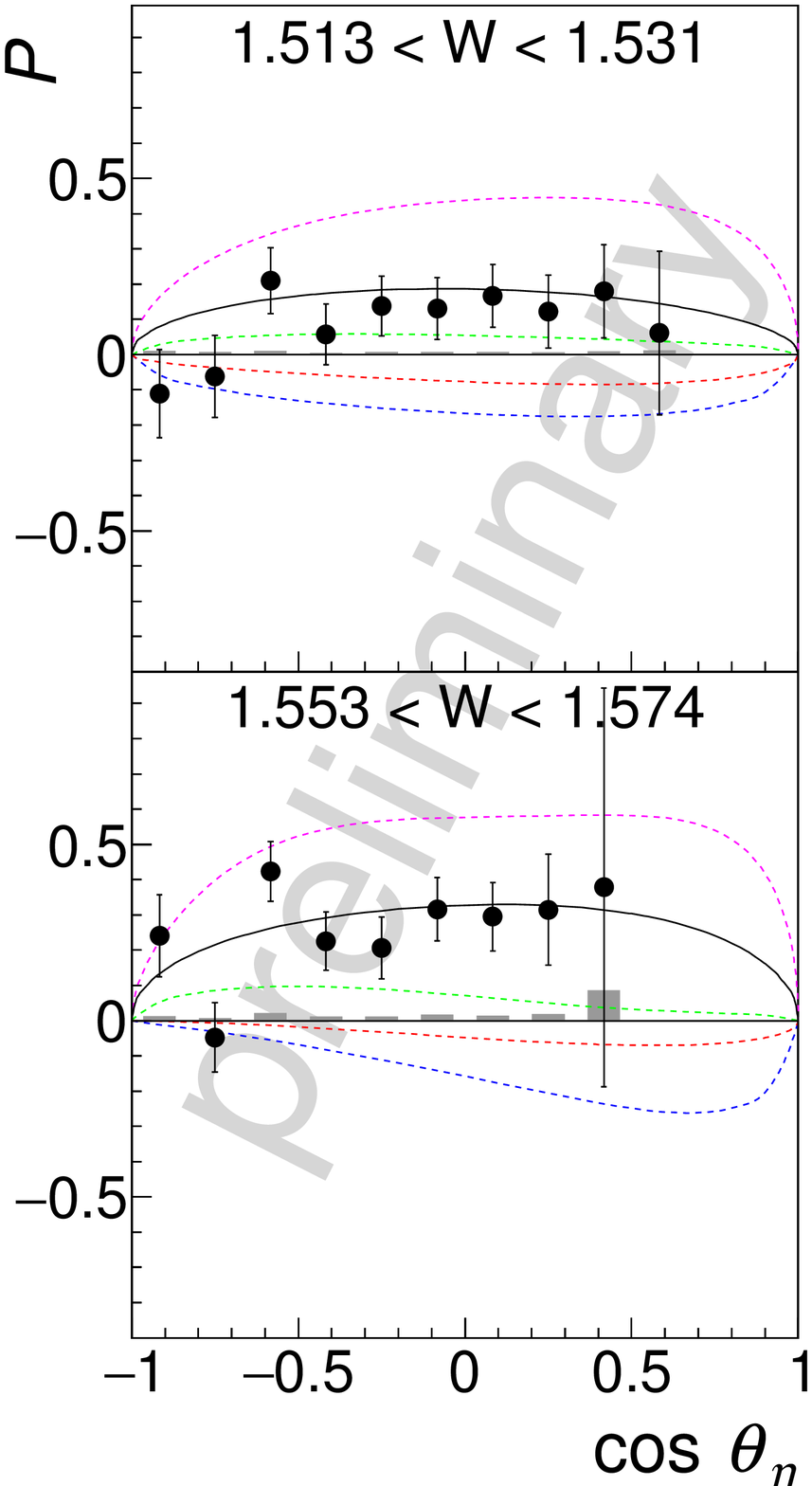}%
	\hfill%
	\includegraphics[height=5.2cm]{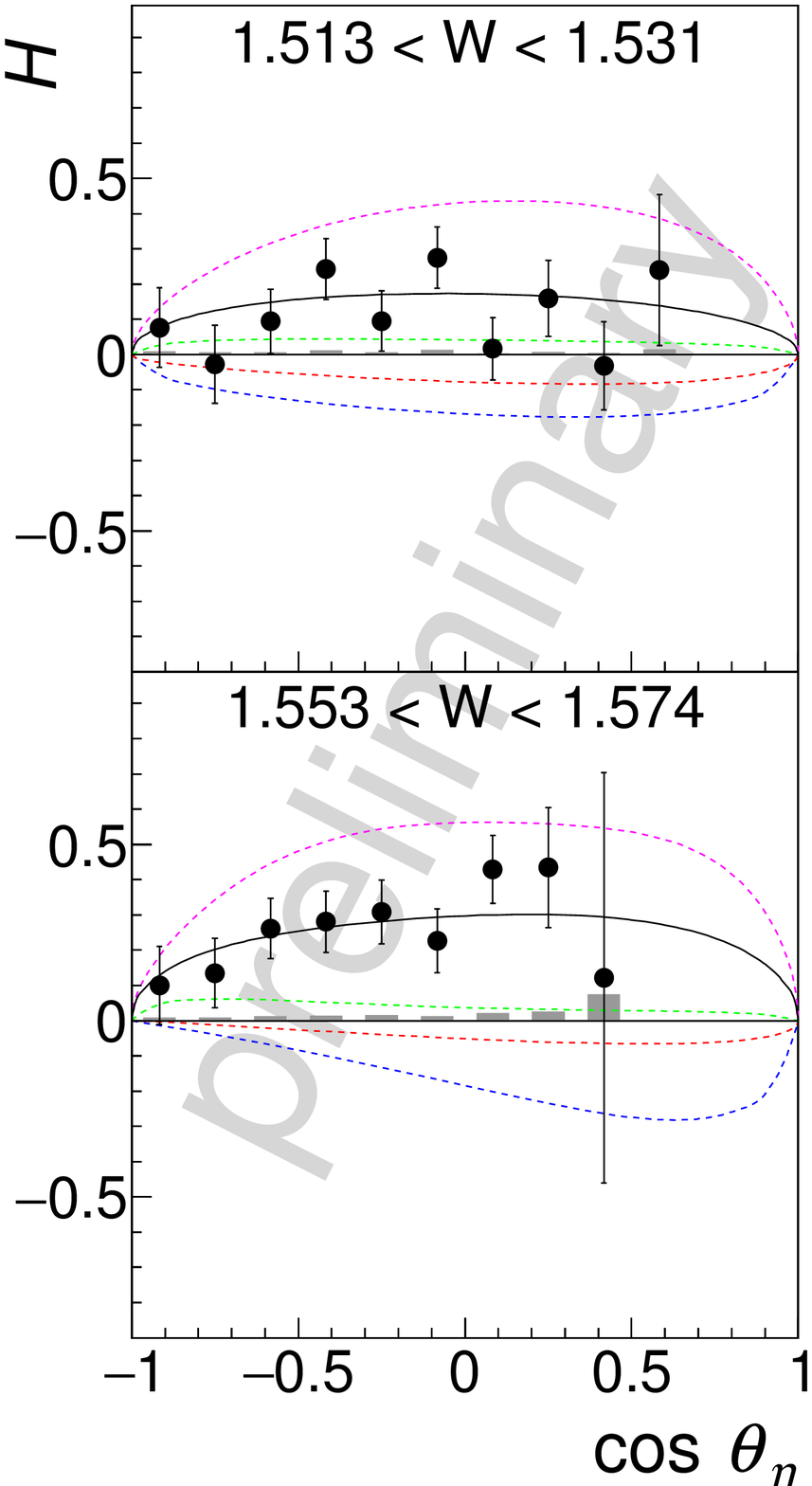}%
	\hfill%
	\vspace*{-2mm}%
	\caption{Preliminary results for the observables $T$, $P$, and $H$ as a function of the scattering angle and the $\gamma p$ invariant mass W (in GeV, only every second bin is shown here). Earlier ELSA data (red) \cite{bock98} and recent MAMI results (green) \cite{akondi14} are shown for comparison. The solid black line represents a preliminary BnGa fit. The data are compared to predictions (dashed curves) from BnGa2011-02 (red) \cite{anisovich12}, MAID (green) \cite{maid07}, SAID GE09 (blue) \cite{mcnicoll10}, and J\"uBo 2015 (magenta) \cite{roenchen15}.}%
	\label{fig:eta_TPH}%
\end{figure}
Preliminary results for the polarisation observables $T$, $P$, and $H$ are shown in Figure~\ref{fig:eta_TPH}.
Large deviations from the data are observed for the predictions from MAID \cite{maid07}, SAID \cite{mcnicoll10}, BnGa2011 \cite{anisovich12}, Gie\ss{}en \cite{shklyar13}, and the J\"uBo model \cite{roenchen15}, emphasising how important these new data are to constrain the amplitudes for $\eta$ photoproduction.

The analysis by the BnGa group of our new data on $T$, $P$, and $H$, together with not yet published data on $E$ and $G$ and further data from Mainz ($T$, $F$) \cite{akondi14} and Jefferson Laboratory \cite{senderovich15} ($E$) is presently ongoing. The results will be published in the near future \cite{mueller16}.

\section{Summary and outlook}
In $\pi^0$ photoproduction, the unprecedented precision of the data significantly reduces the errors of the PWAs, leading to a more precise determination of resonance parameters. In $\eta$ photoproduction, where several observables are now measured for the first time, the new data are crucial to constrain the photoproduction amplitudes. Further reaction channels are also being investigated. In particular multi-meson final states like $p \pi^0 \pi^0$ or $p \pi^0 \eta$ are sensitive to cascade decays of higher-mass resonances via intermediate $N^*$ and $\varDelta^*$ states \cite{sokhoyan15a,sokhoyan15b,mahlberg16}. \medskip

\begin{acknowledgement}%
We thank the technical staff of ELSA, the polarised target group, and the participating institutions for their invaluable contributions to the success of the experiment.
We acknowledge support from the \textit{Deutsche Forschungsgemeinschaft} (SFB/TR16) and \textit{Schweizerischer Nationalfonds}.
\end{acknowledgement}

\bibliography{hartmann}%

\end{document}